\documentclass[prd,twocolumn,showpacs,floatfix,amsmath,nofootinbib,amssymb,floatfix]{revtex4}
\usepackage{graphicx,color,dcolumn,booktabs,bm}
\usepackage{longtable,lscape}
\usepackage{txfonts}
\usepackage{overpic}
\usepackage{amssymb}
\usepackage{indentfirst}
\usepackage{feynmf}   
\usepackage{slashed}  
\usepackage{cases}
\usepackage{color}
\usepackage{multirow}
\usepackage{epstopdf}
\usepackage{graphicx,color,dcolumn,booktabs,bm}
\usepackage[colorlinks, citecolor=blue,anchorcolor=red,menucolor=red, linkcolor=red,filecolor=red,runcolor=red,urlcolor=blue,frenchlinks=red]{hyperref}

\graphicspath{{Figures/}} %

\begin{document}

\title{Understanding $B^-\rightarrow X(3823)K^-$ via rescattering mechanism and predicting $B^-\to \eta_{c2} (^1D_2)/\psi_3(^3D_3)K^-$}

\author{Hao Xu$^{1,2}$}\email{xuh2013@lzu.cn}
\author{Xiang Liu$^{1,2}$\footnote{Corresponding author}}\email{xiangliu@lzu.edu.cn}
\author{Takayuki Matsuki$^{3,4}$}\email{matsuki@tokyo-kasei.ac.jp}
\affiliation{$^1$School of Physical Science and Technology, Lanzhou University,
Lanzhou 730000, China\\
$^2$Research Center for Hadron and CSR Physics,
Lanzhou University $\&$ Institute of Modern Physics of CAS,
Lanzhou 730000, China\\
$^3$Tokyo Kasei University, 1-18-1 Kaga, Itabashi, Tokyo 173-8602, Japan\\
$^4$Theoretical Research Division, Nishina Center, RIKEN, Wako, Saitama 351-0198, Japan}

\begin{abstract}

We study the observed $B^-\rightarrow X(3823)K^-$ decay via rescattering mechanism and show that 
this branching ratio is well reproduced by this mechanism. We further extend this theoretical framework to investigate the decays of $B^-\to\eta_{c2} (^1D_2)/\psi_3(^3D_3)K^-$, where the $\eta_{c2} (^1D_2)$ and the $\psi_3(^3D_3)$ are $D$-wave charmonium partners of the $X(3823)$. Our results show that the branching ratios, $B^-\to \eta_{c2}(^1D_2) K^-$ and $B^-\to \psi_3(^3D_3) K^-$, are of the order of $10^{-5}$, which can be accessible at 
LHCb, Belle and forthcoming BelleII.
\end{abstract}

\pacs{14.40.Pq, 13.25.Hw}

\maketitle

\section{introduction}\label{sec1}

In the past decade, abundant charmonium and charmonium-like states have been discovered. Some of them
 cannot fit into traditional quark model predictions and leave many puzzles. So it attracts great attention to explore
 their inner structure and interaction mechanism (see Refs.~\cite{Liu:2013waa,Chen:2016qju} for a review). An important feature
 is that many $XYZ$ or charmonium states strongly couple to open charmed mesons, and as a result, it leads to a quite
 interesting phenomenon in many processes, which is called a rescattering effect. For example, in a hadronic transition
 process of a charmonium, instead of gluon-emission, the charmonium can first decay into charmed and anti-charmed 
 mesons, and these two mesons rescatter each other into a charmonium plus a light meson. Such a rescattering effect
 has been extensively studied by many authors (see Refs.~\cite{Liu:2006dq,Liu:2006df,Meng:2007cx,Meng:2007tk,Liu:2009iw,Guo:2010zk,Chen:2011qx,Chen:2011zv,Guo:2016iej}).
Their results indicate that rescattering effects can significantly change the line shapes of three-body decays and enhance the results of the Okubo-Zweig-Iizuka (OZI)-suppressed processes. 
  
Another example is given in the situation that the rescattering effect is combined with non-leptonic $B$ meson decays. As we will see later, such a rescattering effect even plays a dominant role. On the other hand, in a naive factorization 
approach which is normally adopted for non-leptonic processes, the amplitudes of some processes such as $B^-\to \chi_{c0}K^-$ vanish (see Sec. \ref{sec2}).
In Ref.~\cite{Colangelo:2002mj}, 
authors explained the large experimental branching fraction of the process $B^-\to\chi_{c0}K^-$ applying the rescattering
mechanism. Later, they also studied $B^-\to h_cK^-$ process using the same mechanism and predicted its branching
ratio \cite{Colangelo:2003sa}. In Ref.~\cite{Cheng:2004ru}, authors systematically studied rescattering effects
 on non-leptonic $B$ meson decays and their impact on direct CP violations. Reference~\cite{Wang:2008hq} studied process
 $B^0\to \eta_c K^*$ using rescattering mechanism and reproduced the experimental data. We also notice that in an earlier 
 time, authors in Refs.~\cite{Li:1996cj,Dai:1999cs} have already applied a rescattering mechanism to study non-leptonic $D$ meson
 decays.

In 2013, Belle observed a new charmonium-like state called $X(3823)$ in the $\chi_{c1} \gamma$ final state in the process $B^-\to
X(3823)K^-$ \cite{Bhardwaj:2013rmw} with measured mass 3823.1 $\pm$1.8(stat.)$\pm$0.7(syst.) MeV and significance 3.8$\sigma$. Recently, BESIII confirmed 
$X(3823)$ in the process $e^+e^- \to \pi^+ \pi^- \gamma \chi_{c1}$ with measured mass 3821.7$\pm$1.3(stat.)$\pm$0.7(syst.) MeV, width less than 16 MeV and significance 6.2$\sigma$ \cite{Ablikim:2015dlj}. $X(3823)$ is expected to be the long missing $\psi_2(1^3D_2)$ with $J^{PC}=2^{--}$.
This is because, first, the mass of $X(3823)$ is consistent with the quark model prediction \cite{Eichten:2002qv,Ebert:2002pp}. Secondly, since the mass 
 of $X(3823)$ is below any open charm threshold ($D \bar{D}$ channel is forbidden by parity conservation),
 the width is quite narrow as expected and as observed. $X(3823)$ largely decays to
 $\chi_{c1}\gamma$, which is the  channel discovered in Belle and BESIII. Furthermore, the upper lmit of the ratio 
 $B(X(3823)\to\chi_{c2}\gamma)/B(X(3823)\to\chi_{c1}\gamma)$ was determined to be $<0.41$ by Belle and $<0.42$ by 
 BESIII, which is consistent with theoretical predictions in Refs.~\cite{Eichten:2002qv,Ebert:2002pp,Qiao:1996ve,Ko:1997rn}.
 Therefore $X(3823)$ is believed to be $\psi_2(^3D_2)$.
 
 In Ref.~\cite{Wang:2015xsa}, authors studied the OZI-suppressed process $X(3823)\to J\psi\pi\pi$ via a rescattering
 effect. Their calculation shows that since the mass of $X(3823)$ is close to the $D \bar D^*$ threshold, a rescattering effect can significantly change the line shape
 of the final $\pi \pi$ mass spectrum. In this work, we will focus on another aspect to investigate the
 resacttering effect on $X(3823)$, i.e., the $X(3823)$ production via a $B$ meson decay. We will illustrate
that the naive factorized amplitude of the process $B^-\to X(3823) K^-$ vanishes, and hence it provides us another good example
 to see how important the scattering effect is.
 
Besides $X(3823)$, there are still missing two other $D$-wave low-lying charmonia, i.e., $\eta_{c2}(^1D_2)$ with $J^{PC}=2^{-+}$
 and $\psi_{2}(^3D_3)$ with $J^{PC}=3^{--}$. Their predicted masses and decay properties are given in Refs.~\cite{Ebert:2002pp,Eichten:2002qv}.
 The naive factorized amplitudes for the processes
 $B^- \to \eta_{c2} K^-$ and $B^- \to \psi_2 K^-$ vanish, for which we will also apply the rescattering mechanism.
Their production rates in the $B$ decay will be a valuable information for experiments.
 
 This paper is organized as follows. After introduction, we study the decay process $B^-\rightarrow X(3823)K^-$ 
 through the rescattering mechanism in Sec.~\ref{sec2}. In Sec.~\ref{sec3}, we make predictions of the production rates for
 the processes $B^- \to \eta_{c2} K^-$ and $B^- \to \psi_2 K^-$. In the final Section, we give discussions and conclusion.

\section{$B^-\rightarrow X(3823)K^-$ via rescattering mechanism} \label{sec2}

First we will show that the naive factorization approach (see Ref. \cite{Buchalla:1995vs}) fails to describe our discussed processes.
When studying $B^-\rightarrow X(3823)K^-$ in this approach, the effective weak Hamiltonian is written as 
\begin{eqnarray}\label{HW}
H_W&=&\frac{G_F}{\sqrt{2}}\Big\{V_{cb}V_{cs}^*\big[c_1(\mu)\mathcal{O}_1(\mu)+c_2(\mu)\mathcal{O}_2(\mu)\big] \nonumber\\
   &&-V_{tb}V_{ts}^*\sum_{i=3}^{10} c_i(\mu)\mathcal{O}_i(\mu)\Big\}+H.c. \;,
\end{eqnarray}
where the operators $\mathcal{O}_i$ read as
\begin{eqnarray}
\mathcal{O}_1&=&(\overline{s}_\alpha b_\beta)_{V-A}(\overline{c}_\beta c_\alpha)_{V-A} \;, \nonumber\\ 
\mathcal{O}_2&=&(\overline{s}_\alpha b_\alpha)_{V-A}(\overline{c}_\beta c_\beta)_{V-A} \;,\nonumber\\
\mathcal{O}_{3(5)}&=&(\overline{s}_\alpha b_\alpha)_{V-A} \sum_{q} (\overline{q}_\beta q_\beta)_{V-A(V+A)} \;,\nonumber\\
\mathcal{O}_{4(6)}&=&(\overline{s}_\alpha b_\beta)_{V-A}\sum_{q}(\overline{q}_\beta q_\alpha)_{V-A(V+A)}  \;,\nonumber\\
\mathcal{O}_{7(9)}&=&\frac{3}{2}(\overline{s}_\alpha b_\alpha)_{V-A} \sum_{q}e_q(\overline{q}_\beta q_\beta)_{V+A(V-A)} \; ,\nonumber\\
\mathcal{O}_{8(10)}&=&\frac{3}{2}(\overline{s}_\alpha b_\beta)_{V-A} \sum_{q}e_q(\overline{q}_\beta q_\alpha){_{V+A(V-A)}} \; . \nonumber
\end{eqnarray}
Thus, the factorized amplitude of the process $B^-\rightarrow X(3823)K^-$ can be expressed as
\begin{eqnarray} \label{fm}
&&\mathcal{M}(B^-\rightarrow X(3823)K^-) \nonumber\\
&&=\frac{G_F}{\sqrt{2}}V_{cb}V^*_{cs}\left(a_2(\mu)+\sum_{i=3,5,7,9}a_i(\mu)\right) \langle K^-|(\overline{s}b)_{V-A}|B^-\rangle \nonumber\\
&&\quad\times\langle X(3823)|(\overline{c}c)_{V\mp A}|0\rangle \;
\end{eqnarray}
with $a_2=c_2+c_1/N_c$ and $a_i=c_i+c_{i+1}/N_c$. In this work, $X(3823)$ is treated as a $D$-wave charmonium with quantum numbers $J^{PC}=2^{--}$. When checking the factorized amplitude in Eq.~(\ref{fm}), we find the matrix element $\langle X(3823)|(\overline{c}c)_{V\mp A}|0\rangle=0$ {due to the Lorentz invariance.  Hence this leads to vanishing of the branching ratio of $B^-\rightarrow X(3823)K^-$ in the naive factorization approach}. 

However, the Belle measurement~\cite{Bhardwaj:2013rmw} shows combined branching fraction $\mathcal{BR}(B^-\to X(3823) K^-)\times \mathcal{BR}(X(3823)\to \chi_{c1} \gamma)=(9.7\pm2.8\pm1.1) \times 10^{-6}$. To obtain the value of $\mathcal{BR}(B^-\to X(3823) K^-)$, we consider  the theoretical partial widths of $X(3823)$ decaying into $\chi_{c1} \gamma$, $\chi_{c2} \gamma$, $ggg$, and $J/\psi \pi \pi$ which are given by $\Gamma(X(3823)\to \chi_{c1} \gamma)=215$ keV \cite{Ebert:2002pp}, 
 $\Gamma(X(3823)\to \chi_{c2} \gamma)=59$ keV \cite{Ebert:2002pp}, $\Gamma(X(3823)\to ggg)=36$ keV \cite{Eichten:2002qv} and 
$\Gamma(X(3823)\to J\psi \pi \pi) \simeq 160$ keV \cite{Wang:2015xsa}, respectively. Summing up all the above partial widths, we can roughly estimate the total decay width of $X(3823)$ to be 470 keV, with which we get $\mathcal{BR}(X(3823)\to \chi_{c1} \gamma)=46\%$. Then, we can extract {
 \begin{eqnarray} \label{BRBtoX3823K}
 \mathcal{BR}(B^-\to X(3823) K^-)= (2.10\pm0.65)\times 10^{-5}  \; ,
 \end{eqnarray}
where the error comes from the combined branching fraction of the Belle measurement. }It shows that there exists a non-zero contribution to the $B^-\to X(3823) K^-$ decay. 

\begin{figure}[htpb]
	\begin{center}
		\includegraphics[scale=0.45]{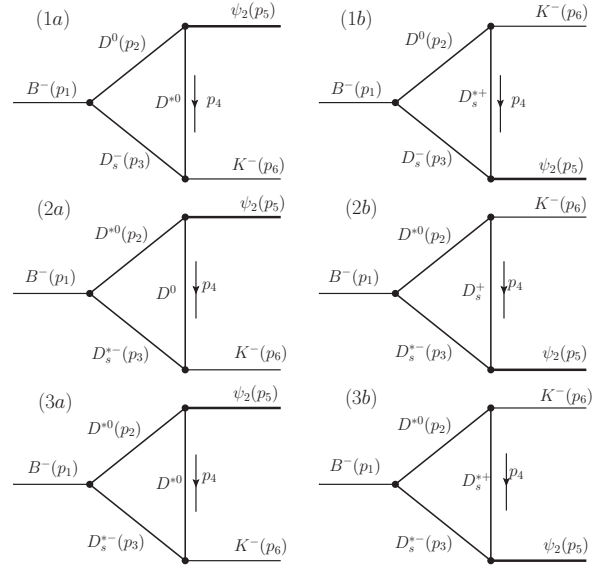}
		\caption{(color online). The schematic diagrams for depicting the $B^-\rightarrow X(3823)K^-$ decay via the rescattering mechanism. Note that 
			$\psi_2$ denotes $X(3823)$.}\label{hadronloopsof3823}
	\end{center}
\end{figure}

To understand the discrepancy between the experimental data and theoretical estimate from the naive factorization approach, we study $B^-\to X(3823) K^-$ by introducing the rescattering mechanism, which was proposed in   
Ref.~\cite{Colangelo:2002mj}. They indicated that such a nonleptonic process should have a large nonfactorizable 
 contribution that comes from the rescattering mechanism. For the discussed $B^-\to X(3823) K^-$ process, $B^-$ first decays into intermediate charmed and anti-charmed meson pair, and then they transit into final states, $X(3823)$ and $K^-$. The typical diagram describing the rescattering effect on $B^-\to X(3823) K^-$ can be found in Fig.~\ref{hadronloopsof3823}. In the following, we calculate these rescattering processes of $B^-\to X(3823) K^-$ to test whether the extracted branching ratio given by Eq. (\ref{BRBtoX3823K}) can be understood under the rescattering mechanism. 

In order to calculate these triangle diagrams at hadron level, we need to introduce the effective Lagrangians  corresponding to each interaction vertex. {As for the weak vertex $B^-\rightarrow D^{(*)0} D^{(*)-}_s$, we also assume the naive factorization of the amplitude. Neglecting the small contributions from the operators $\mathcal{O}_3\sim\mathcal{O}_{10}$ in Eq.~(\ref{HW}), the transition matrix element can be factorized as
\begin{eqnarray} \label{BtoDD}
\langle D^{(*)0} D^{(*)-}_s |H_W| B^- \rangle&=&\frac{G_F}{\sqrt{2}} V_{cb} V_{cs}^* a_1 \langle D^{(*)0}|V^\mu-A^\mu|B^-\rangle \nonumber \\
 &&\langle D^{(*)-}_s|V_\mu-A_\mu|0\rangle, 
\end{eqnarray}
where $a_1=c_1+c_2/N_c$.
One should notice that this naive factorization for the process $B^-\rightarrow D^{(*)0} D^{(*)-}_s$ had been shown to be a good approximation in Ref. \cite{Luo:2001mc}. 
The matrix element appearing in Eq.~(\ref{BtoDD}) can be simply written in terms of form factors and decay constants.}

 we use the following matrix elements that contain only one form factor $\xi$, i.e., Isgur-Wise function \cite{Cheng:2003sm}:
\begin{eqnarray}
\langle D^0(v^\prime)|V^\mu|B^-(v)\rangle&=&\sqrt{m_B m_D}\xi(v\cdot v^\prime)(v^\prime+v)^\mu, \nonumber\\
\langle D^{*0}(v^\prime,\epsilon)|V_\mu|B^-(v)\rangle&=& i\sqrt{m_Bm_{D^*}}\xi(v\cdot v^\prime)\varepsilon_{\mu\nu\alpha\beta}\epsilon^{\nu *}v^{\prime\alpha}v^\beta, \nonumber\\
\langle D^{*0}(v^\prime,\epsilon)|A^\mu|B^-(v)\rangle&=& \sqrt{m_Bm_{D^*}}\xi(v\cdot v^\prime)\Bigl((1+v\cdot v^\prime)g^{\alpha \mu} \nonumber\\
&&-v^\alpha v^{\prime\mu}\Bigr) \epsilon^*_\alpha, \nonumber\\
\langle 0|A^\mu|D_s(v)\rangle&=& f_{D_s}m_{D_s}v^\mu, \nonumber\\
\langle0|V^\mu|D^*_s(v,\epsilon)\rangle&=&f_{D^*_s}m_{D^*_s}\epsilon^\mu.  \nonumber
\end{eqnarray}
Using these matrix elements, one further obtains the transition amplitudes:
\begin{eqnarray}
&&\langle D^0(p_2) D^-_s(p_3) | B^-(v_1) \rangle  \nonumber \\
 &&= \frac{G_F}{\sqrt{2}} V_{cb} V^*_{cs} a_1 \sqrt{m_1 m_2} \xi(v \cdot v^\prime) \left(  \frac{p^\mu_2}{m_2} + v^\mu_1\right)  f_3 p_{3\mu} \; ,\\
&&\langle D^{*0}(p_2) D^{*-}_s(p_3) | B^-(v_1) \rangle  \nonumber \\
&&= \frac{G_F}{\sqrt{2}} V_{cb} V^*_{cs} a_1 \sqrt{m_1 m_2} \xi(v \cdot v^\prime) \Bigg( i \epsilon_{\mu\nu\alpha\beta} \frac{p^\alpha_2}{m_2}v^\beta_1-(1+\omega)g_{\nu\mu}  \nonumber \\
&& \quad +v_{1\nu}\frac{p_{2\mu}}{m_2} \Bigg) \epsilon^{*\nu}_2 f_3 m_3 \epsilon^{*\mu}_3 \; , \\
&&\langle D^0(p_2) D^{*-}_s(p_3) | B^-(v_1) \rangle  \nonumber \\
&&= \frac{G_F}{\sqrt{2}} V_{cb} V^*_{cs} a_1 \sqrt{m_1 m_2} \xi(v \cdot v^\prime) \left(  \frac{p^\mu_2}{m_2} + v^\mu_1\right)  f_3 m_3 \epsilon^*_{3\mu} \;,\\
&&\langle D^{*0}(p_2) D^{-}_s(p_3) | B^-(v_1) \rangle  \nonumber \\
&&= \frac{G_F}{\sqrt{2}} V_{cb} V^*_{cs} a_1 \sqrt{m_1 m_2} \xi(v \cdot v^\prime) \Bigg( i \epsilon_{\mu\nu\alpha\beta} \frac{p^\alpha_2}{m_2}v^\beta_1-(1+\omega)g_{\nu\mu}  \nonumber \\
&& \quad +v_{1\nu}\frac{p_{2\mu}}{m_2} \Bigg) \epsilon^{*\nu}_2 f_3 p^{\mu}_3 \; , 
\end{eqnarray}
where { $m_1$ and $m_2$ are the masses of $B^-$ and $D^{(*)0}$, respectively, $f_3$ is a decay constant of the particle carrying a momentum $p_3$, and $a_1=c_1+c_2/N_c$ as defined in Eq. (\ref{BtoDD}).}

For the $D^{(*)}_s D^{(*)} K$ interactions, we adopt the effective Lagrangians respecting both the heavy quark symmetry and chiral symmetry. For a heavy-light
meson system,  there exist heavy quark spin symmetry and heavy quark
flavor symmetry \cite{Neubert:1993mb} in the heavy quark limit $m_Q\to \infty$. As a consequence, heavy-light mesons {are degenerate and are classified into} different multiplets,
such as an $H$ doublet $(0^-,1^-)$ with quantum number of light degrees of freedom $j_\ell^P=\frac{1}{2}^-$. The multiplet can be described
 by an effective hadron field respecting the heavy quark symmetry. For example, the field of an $H$ doublet $(D,D^*)$ or $(D_s,D^*_s)$ is given by
\begin{eqnarray}\label{Hfield}
H_a=\left( \frac{1 + \slashed{v}}{2} \right) ( D^{*\mu}_a \gamma_\mu + i D_a \gamma_5 ) \; 
\end{eqnarray}
with $a$ the flavor index and $v$ the meson velocity, where the fields $ D^{*\mu}_a $ and $ D_a $ contain a
normalization factor $\sqrt{m_M}$ and have dimension $3/2$.

For an $H$ field coupled with an octet chiral multiplet, the effective Lagrangian reads \cite{Wise:1992hn}:
\begin{eqnarray}\label{lagrangianHH0-}
\mathcal {L}_H=i g_H \textnormal{Tr} \left[ H_b \gamma_\mu \gamma_5 \mathcal{A}^\mu_{ba} \bar{H}_a \right],
\end{eqnarray}
where $ \mathcal{A}^\mu_{ba} =i/f_\pi \partial^\mu M_{ba}+...$ with $M_{ba}$ the octet of light pseudoscalar mesons
and $f_\pi=132$ MeV. $\bar{H_a}$ satisfies the relation $\bar{H_a}=\gamma^0 H^\dagger_a \gamma^0$. 
By expanding the Lagrangian in Eq.~(\ref{lagrangianHH0-}), effective Lagrangians for the vertexes
$D^{(*)}_s D^{(*)} K$ are explicitly given by
 \begin{eqnarray}
\mathcal {L}_{D D_{s}^* K } &=& i g_{DD^*_sK} \bar{D}^*_{s\mu} D \partial^\mu K \; , \\
\mathcal {L}_{D_s D^* K } &=& -i g_{D_sD^*K} \bar{D}_{s} D^*_\mu \partial^\mu K \; , \\
\mathcal {L}_{D^*_s D^* K } &=& - g_{D^*_sD^*K} \epsilon_{\mu\nu\alpha\beta} \partial^\mu \bar{D}^{*\nu}_s \partial^\alpha D^{*\beta} K \;,
 \end{eqnarray}
where the coupling constants are related to $g_H$ as,
 \begin{eqnarray}
 g_{DD^*_sK} &=& g_{D_sD^*K}=\sqrt{m_{D^*_s} m_D} \frac{2 g_H}{f_\pi} \; , \\
 g_{D^*_sD^*K} &=& \frac{\sqrt{m_{D^*_s} m_{D^*}}}{m_{D^*}} \frac{2 g_H}{f_\pi} \; .
 \end{eqnarray}
 
The vertexes $X(3823)D^{(*)}D^*$ and $X(3823)D^{(*)}_sD^*_s$ are additionally involved in our calculation, for which we also use the effective Lagrangians respecting the heavy quark symmetry. However, for a charmonium system, the heavy
 quark flavor symmetry does not hold, where only the heavy quark spin symmetry remains \cite{Casalbuoni:1996pg}.
 Thus, charmonia with the same orbital angular momentum $L$ but with different total spin can form a multiplet.
 In our case, $X(3823)$ belongs to a $D$-wave multiplet \cite{Casalbuoni:1996pg,Casalbuoni:1992fd}, which is defined by
 \begin{eqnarray}
X^{\mu\nu} 
 &&= {\frac{1+\slashed{v}}{2}} \bigg\{  \psi^{\mu\nu\alpha}_3 \gamma_\alpha + \frac{1}{\sqrt{6}} \Big( \epsilon^{\mu\delta\alpha\beta} v_\delta \gamma_\alpha \psi^\nu_{2\beta} + \epsilon^{\nu\delta\alpha\beta} v_\delta \gamma_\alpha \psi^\mu_{2\beta}\Big) \nonumber \\
 &&\quad + \frac{\sqrt{15}}{10} \Big( (\gamma^\mu-v^\mu) \psi^\nu + \psi^\mu (\gamma^\nu-v^\nu) \Big)  \nonumber \\
 && \quad- \frac{1}{\sqrt{15}} ( g^{\mu\nu} - v^\mu v^\nu ) \gamma_\alpha \psi^\alpha + \eta^{\mu\nu}_2 \gamma_5 \bigg\} {\frac{1-\slashed{v}}{2}} \;.
 \end{eqnarray}
In the above expression,  the fields $\psi_3$, $\psi_2$, $\psi$, and $\eta_2$ denote the charmonia with quantum numbers $J^{PC}=3^{--}$, $2^{--}$, $1^{--}$ and $2^{-+}$, respectively, where $\psi_2$ corresponds to the discussed $X(3823)$.

 For the coupling of $D$-wave chamonium multiplet with charmed mesons, their effective Lagrangian reads \cite{Wang:2015xsa}
\begin{eqnarray} \label{lagrangianXHH}
\mathcal{L}_X = g_X \textnormal{Tr} \bigg[ X^{\mu\nu} \bar{H}_{\bar{Q}a} ( \overrightarrow{\partial}_\mu - \overleftarrow{\partial}_\mu ) \gamma_\nu \bar{H}_{Qa} \bigg] \;,
\end{eqnarray}
where $H_{Qa}$ is given by Eq.~(\ref{Hfield}), and $H_{\bar{Q}a}$ is
\begin{eqnarray}
H_{\bar{Q}a} = ( D^{*\mu}_a \gamma_\mu + i D_a \gamma_5 ) \left( \frac{1 - \slashed{v}}{2} \right) \; ,
\end{eqnarray}
which is obtained by the charge conjugation transformation. The fields $\bar{H}_{\bar{Q}a}$ and $\bar{H}_{Qa}$ 
appearing in Eq.~(\ref{lagrangianXHH}) are defined as $\bar{H}_{\bar{Q}a}=\gamma^0 H^\dagger_{\bar{Q}a} \gamma^0$
 and $\bar{H}_{Qa}=\gamma^0 H^\dagger_{Qa} \gamma^0$, respetively.
Then, the explicit forms of the  $X(3823)D^{(*)}D^*$ ($\psi_2D^{(*)}D^*$)  interactions can be obtained as
\begin{eqnarray}
\mathcal{L}_{\psi_2DD^*} 
 &=& g_{\psi_2DD^*}  \psi^{\mu\nu}_2 (\partial_\nu \bar{D} D^*_\mu-\bar{D} \partial_\nu D^*_\mu) \nonumber \\
 &&+g_{\psi_2DD^*} \psi^{\mu\nu}_2 (\bar{D}^*_\mu \partial_\nu D - \partial_\nu \bar{D}^*_\mu D)  \; , \label{lagrangianX3823DDstar}\\
 \mathcal{L}_{\psi_2D^*D^*} 
 &=& i g_{\psi_2D^*D^*} \varepsilon_{\mu\nu\alpha\beta} \partial^\mu \psi^{\nu\rho}_2 D^*_\rho \partial^\alpha \bar{D}^{*\beta} \nonumber \\
 && + i g_{\psi_2D^*D^*} \varepsilon_{\mu\nu\alpha\beta} \partial^\mu \psi^{\nu\rho}_2 \partial^\alpha  \bar{D}^*_\rho D^{*\beta}  \; , \label{lagrangianX3823DstarDstar}
\end{eqnarray}
where 
\begin{eqnarray}
g_{\psi_2DD^*}&=&\sqrt{6} g_X \sqrt{m_D m_{D^*} m_{\psi_2}} \;,  \\
g_{\psi_2D^*D^*}&=&\frac{-4 g_X}{\sqrt{6}}  \frac{\sqrt{m_{D^*} m_{D^*}m_{\psi_2}} }{m_{\psi_2}} \;.
\end{eqnarray}
The Lagrangians of vertexes $\psi_2D^{(*)}_sD^*_s$ are similar to those shown in Eqs.~(\ref{lagrangianX3823DDstar})-(\ref{lagrangianX3823DstarDstar}),
 where the corresponding coupling constants satisfy $g_{\psi_2D_sD^*_s}=g_{\psi_2DD^*}$ and $g_{\psi_2D^*_sD^*_s}=g_{\psi_2D^*D^*}$ if the SU(3) flavor symmetry holds.

Applying the Cutkosky cutting rule \cite{Peskin:1995ev}, the imaginary parts of the decay amplitudes of $B^-\to X(3823) K^-$
can be obtained, {for example, for the amplitude of the diagram Fig.~\ref{hadronloopsof3823}~(1a) as} 
\begin{eqnarray}
&& Abs^{B^-\to\psi_2 K^-}_{(1a)}  \nonumber \\
&&= \frac{|p_2|}{32 \pi^2 m_1} \int d\Omega \frac{G_F}{\sqrt{2}} V_{cb} V^*_{cs} a_1 \sqrt{m_1 m_2} \xi(\omega) \left(v_1^\gamma + \frac{p_2^\gamma}{m_2} \right)
f_3 p_{3\gamma}  \nonumber \\
&& \quad\times  (-i) g_{\psi_2DD^*} \epsilon^{*\mu\nu}_5 (p_{2\nu}+p_{4\nu}) \left( -g_{\mu\alpha} + \frac{p_{4\mu}p_{4\alpha}}{m^2_4} \right) 
g_{D_sD^*K} p^\alpha_6   \nonumber \\
&& \quad\times \frac{1}{p^2_4-m^2_4} \mathfrak{F}^2(p_4^2) \; .
\end{eqnarray}

Here $m_i$ ($i=1, 2, 3, ...$) denotes the mass of the particle carrying momentum $p_i$ in Fig.~\ref{hadronloopsof3823}, and $\omega=v\cdot v^\prime$.
{ Other amplitudes are given in Appendix}. On the other hand, we need also introduce form factors to 
compensate the off-shell effect of the exchanged $D^{(*)}_{(s)}$ in Fig.~\ref{hadronloopsof3823}. The concrete expression of the form factor is \cite{Liu:2006df,Cheng:2004ru}
\begin{eqnarray}\label{ff}
\mathfrak{F}(q^2)=\frac{\Lambda^2-m^2}{\Lambda^2 - q^2} \; ,
\end{eqnarray}
where the cutoff parameter $\Lambda$ can be parameterized as
\begin{eqnarray}\label{cut}
\Lambda = m+ {\alpha} \Lambda_{QCD}
\end{eqnarray}
with $\Lambda_{QCD} = 220$ MeV. $m$ denotes the mass of the exchanged meson. 

The total absorptive part of the amplitude of the process $B^-\to X(3823) K^-$ is 
\begin{eqnarray*}
 Abs \left(\mathcal{M}[B^-\to X(3823) K^-] \right)= \sum_{i=1a,...,3b} Abs^{B^-\to\psi_2 K^-}_{(i)} \; ,
\end{eqnarray*}
with which we can estimate the decay width of the process $B^-\to X(3823) K^-$ as
\begin{eqnarray}
\Gamma(B^-\to X(3823) K^-)=\frac{1}{8\pi} \frac{|\vec{p}|}{m^2_{B}} |Abs(\mathcal{M})|^2 \; .
\end{eqnarray}
Here, $\vec p$ denotes the three-momentum of final states in the center of mass frame of $B^-$ meson, and $m_B$ is the mass of $B$ meson. 

{In principle, we may include the real part (dispersive part) of the scattering amplitude through the absorptive part:
\begin{eqnarray}
     Dis(\mathcal{M}(m^2_B))=\frac{1}{\pi} \int\limits_{s}^{+\infty} \frac{Abs(\mathcal{M}(s^\prime))}{s^\prime-m^2_b} ds^\prime.
\end{eqnarray}	
However, as discussed in Ref.~\cite{Cheng:2004ru}, this real part has large uncertainties that come from a newly introduced cut-off parameter and 
 integration itself. Furthermore, since the mass of the $B$ meson is far from the $D$ meson pair threshold, the imaginary part can largely increase
 and become dominant in full amplitude. Hence we assume the absorptive part is dominant as in Ref.~\cite{Cheng:2004ru}, and ignore the 
 dispersive part. }

In order to obtain the results, the values of various parameters should be specified, which include the weak Fermi coupling constant
 $G_F=1.16638\times10^{-5} \textnormal{ GeV}^{-2}$, $V_{cb}=0.04$ and $V_{cs}=1.0$ \cite{Agashe:2014kda}, decay 
 constants $f_{D_s}=f_{D^*_s}=0.24$ GeV and Wilson coefficient $a_1=1.0$ \cite{Colangelo:2003sa}. As for the mass of  $X(3823)$, we adopt the BESIII's result $m_{X(3823)}=3.8217$ 
 GeV \cite{Ablikim:2015dlj} as an input. The strong coupling constants $g_H\simeq0.57$ and $g_X\simeq1.4 \textnormal{ GeV}^{-3/2}$ are given in Ref. \cite{Wang:2015xsa}.
As for the Isgur-wise function, we adopt the form calculated in Ref.~\cite{Cheng:2003sm}:
 \begin{eqnarray} \label{IWfunction}
 \xi(\omega)=1-1.22(\omega-1)+0.85(\omega-1)^2 \;.
 \end{eqnarray}

So far in our calculation the only unknown parameter left is $\alpha$ in Eq.~(\ref{cut}).
{The rescattering mechanism becomes soft in the case of the B meson decay because of heaviness of B meson mass. Since the rescattering mechanism as a long-distant contribution plays an important role to understand $B^-\to X(3823) K^-$, we try to reproduce the experimental branching ratio of $B^-\to X(3823) K^-$ shown in Eq.~(\ref{BRBtoX3823K}) by varying the parameter $\alpha$ to obtain { $\alpha= 0.70\pm0.05$,
	where the error comes from Eq.~(\ref{BRBtoX3823K})}.}
It is obvious that this is not the end of the whole story. 
This value of $\alpha$ can be applied to study similar processes like the productions of $\eta_{c2} (^1D_2)/\psi_3(^3D_3)$ plus a kaon via $B$ meson decays, where $\eta_{c2} (^1D_2)$ and $\psi_3(^3D_3)$ are as the D-wave charmonium partners of $X(3823)$.  In the next section, we illustrate the details of the corresponding deduction.

\section{prediction of $B^-\to \eta_{c2}(^1D_2) K^-$ and $B^-\to \psi_3(^3D_3) K^-$}\label{sec3}
After discussing the $B^-\to X(3823) K^-$ decay, in this section we further investigate the productions of 
two D-wave charmonia $\eta_{c2}(^1D_2)$ with $J^{PC}=2^{-+}$ and $\psi_3(^3D_3)$ with $J^{PC}=3^{--}$ through similar $B$ decay processes. Here, $\eta_{c2}(^1D_2)$ and $\psi_3(^3D_3)$ have not yet been observed in experiment, which also stimulates us to predict the production rates of $B^-\to \eta_{c2} K^-$ and $B^-\to \psi_3 K^-$. 

Similar to the process $B^-\to X(3823) K^-$, the processes $B^-\to \eta_{c2} K^-$ and
 $B^-\to \psi_3 K^-$ are also forbidden if simply considering the naive factorization contribution, since 
$\langle \eta_{c2}(2^{-+})|(\overline{c}c)_{V\mp A}|0\rangle=0$ and $\langle \psi_3(3^{--})|(\overline{c}c)_{V\mp A}|0\rangle=0$. According to the former experience of study of $B^-\to X(3823) K^-$, we need to introduce the rescattering mechanism to estimate the decay rates of these two processes. 

In order to calculate the processes $B^-\to \eta_{c2}(^1D_2) K^-$
 and $B^-\to \psi_3(^3D_3) K^-$, one needs to have the effective Lagrangian
given in Eq.~(\ref{lagrangianXHH}). For the vertexes $\eta_{c2}DD^*$ and $\eta_{c2}D^*D^*$, the corresponding Lagrangians read
\begin{eqnarray}
\mathcal{L}_{\eta_{c2}DD^*} &=& i g_{\eta_{c2}DD^*} \eta_{c2}^{\mu\nu} ( \partial_\nu \bar{D} D^*_\mu - \bar{D} \partial_\nu D^*_\mu ) \nonumber \\
&& - i g_{\eta_{c2}DD^*} \eta^{\mu\nu}_{c2} ( \bar{D}^*_\mu \partial_\nu D - \partial_\nu \bar{D}^*_\mu D)  \; , \label{lagrangianeta2DDstar} \\
\mathcal{L}_{\eta_{c2}D^*D^*} &=& g_{\eta_{c2}D^*D^*} \varepsilon_{\mu\nu\alpha\beta} \partial^\mu \eta_{c2}^{\rho\nu} D^{*\alpha}
\partial_\rho \bar{D}^{*\beta}    \; \label{lagrangianeta2DstarDstar}
\end{eqnarray}
with
\begin{eqnarray}
g_{\eta_{c2}DD^*}&=& 2 g_X \sqrt{m_D m_{D^*} m_{\eta_{c2}}}   \\
g_{\eta_{c2}D^*D^*}&=& 4 g_X \frac{\sqrt{m_{D^*} m_{D^*} m_{\eta_{c2}} } } {m_{\eta_{c2}}}  \; .
\end{eqnarray}

For the vertex $\psi_3D^*D^*$, the Lagrangian is
\begin{eqnarray}
\mathcal{L}_{\psi_3D^*D^*} = g_{\psi_3D^*D^*} \psi^{\mu\nu\alpha}_3 ( \partial_\mu \bar{D}^*_\nu D^*_\alpha - \bar{D}^*_\nu \partial_\mu D^*_\alpha ) \;  \label{lagrangianX3DstarDstar}
\end{eqnarray}
with
\begin{eqnarray}
g_{\psi_3D^*D^*}=4g_X \sqrt{m_{D^*}m_{D^*}m_{\psi_3}} \;.
\end{eqnarray}
The Lagrangians for the vertexes $\eta_{c2}D^{(*)}_sD^*_s$ and $\psi_3D_s^*D_s^*$  have the same form as those shown in Eqs.~(\ref{lagrangianeta2DDstar}), (\ref{lagrangianeta2DstarDstar}), and (\ref{lagrangianX3DstarDstar}),
where we only need to have the relations among the involved coupling constants, i.e., $g_{\eta_{c2}D_sD^*_s}=g_{\eta_{c2}DD^*}$, $g_{\eta_{c2}D^*_sD^*_s}=g_{\eta_{c2}D^*D^*}$ and
$g_{\psi_3D^*_sD^*_s}=g_{\psi_3D^*D^*}$, which are obtained by assuming  
the $SU(3)$ flavor symmetry.

For $B^-\to \eta_{c2} K^-$, the corresponding diagrams are the same as those of $B^-\to X(3823) K^-$, where we only need to make a replacement $X(3823)\to \eta_{c2}$ in the diagrams shown in Fig.~\ref{hadronloopsof3823}. 
With the above preparation, the  absorptive parts of the amplitudes in the process $B^-\to \eta_{c2} K^-$  
can be {obtained, for example, for the amplitude of the diagram Fig.~\ref{hadronloopsof3823}~(1a) as}
\begin{eqnarray}
&& Abs^{B^-\to \eta_{c2} K^-}_{(1a)}  \nonumber \\
&&= \frac{|p_2|}{32 \pi^2 m_1} \int d\Omega \frac{G_F}{\sqrt{2}} V_{cb} V^*_{cs} a_1 \sqrt{m_1 m_2} \xi(\omega) \left(v_1^\gamma + \frac{p_2^\gamma}{m_2} \right)
f_3 p_{3\gamma}  \nonumber \\
&&\quad \times  (-1) g_{\eta_{c2}DD^*} \epsilon^{*\mu\nu}_5 (p_{2\nu}+p_{4\nu}) \left( -g_{\mu\alpha} + \frac{p_{4\mu}p_{4\alpha}}{m^2_4} \right) 
g_{D_sD^*K} p^\alpha_6   \nonumber \\
&&
\quad \times \frac{1}{p^2_4-m^2_4} \mathfrak{F}^2(p_4^2) \; ,
\end{eqnarray}
where $m_i$ ($i=1, 2, 3, ...$) denotes the mass of the particle carrying momentum $p_i$ in Fig.~\ref{hadronloopsof3823}.
{The rest of the amplitudes are given in Appendix.}

The total absorptive part of the amplitude of the process $B^-\to \eta_{c2} K^-$ is 
\begin{eqnarray*}
Abs \left(\mathcal{M}[B^-\to \eta_{c2} K^-] \right) &=&  \sum_{i=1a,...,3b} Abs^{B^-\to \eta_{c2} K^-}_{(i)} \; , 
\end{eqnarray*}

\begin{figure}[htpb]
	\begin{center}
		\includegraphics[scale=0.45]{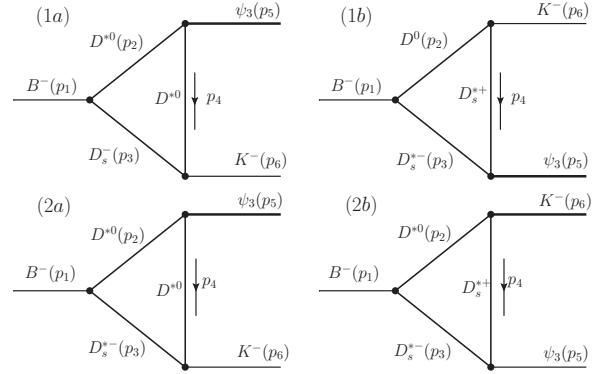}
		\caption{(color online). The schematic diagrams for the decay
			$B^-\rightarrow \psi_3K^-$ via the rescattering mechanism.}\label{FeynmannDiagramX3}
	\end{center}
\end{figure}

For $B^-\to \psi_{3} K^-$, the relevant diagrams are shown in Fig.~\ref{FeynmannDiagramX3}.
The  absorptive parts of the amplitudes of the process $B^-\to \psi_3 K^-$  
can be { obtained, for example, for the amplitude of the diagram Fig.~\ref{FeynmannDiagramX3}~(1a) as}
\begin{eqnarray}
&& Abs^{B^-\to \psi_3 K^-}_{(1a)}  \nonumber \\
&&= \frac{|p_2|}{32 \pi^2 m_1} \int d\Omega \frac{G_F}{\sqrt{2}} V_{cb} V^*_{cs} a_1 \sqrt{m_1 m_2} \xi(\omega)
\Bigg( i\varepsilon_{\gamma\delta\eta\beta} \frac{p^\eta_2}{m_2}v^\beta_1  \nonumber \\
&&\quad - (1+\omega)g_{\delta\gamma} + v_{1\delta} \frac{p_{2\gamma}}{m_2} \Bigg) f_3 p^\gamma_3 \left( -g^\delta_\alpha + \frac{p^\delta_2 p_{2\alpha}}{m^2_2} \right)
i g_{\psi_3D^*D^*}   \nonumber \\
&&\quad \times \epsilon^{*\mu\nu\alpha}_5 (p_{4\mu}+p_{2\mu}) \left( -g_{\nu\theta} + \frac{p_{4\nu} p_{4\theta}}{m^2_4} \right) g_{D^*D_sK} p_{6}^\theta 
\frac{1}{p^2_4-m^2_4}  \nonumber \\
&&\quad \times \mathfrak{F}^2(p^2_4)  \; .
\end{eqnarray}
Note that $m_i$ ($i=1, 2, 3, ...$) denotes the mass of the particle carrying momentum $p_i$ in Fig.~\ref{FeynmannDiagramX3}.
{The rest of the amplitudes are given in Appendix}.

The total absorptive part of the amplitude of the process $B^-\to \psi_3 K^-$ is 
\begin{eqnarray*}
Abs \left(\mathcal{M}[B^-\to \psi_3 K^-] \right)  &=&\sum_{i=1a,...,2b} Abs^{B^-\to \psi_3 K^-}_{(i)} \; .
\end{eqnarray*}

Other input parameters are the masses of two unobserved charmonia $\eta_{c2}$ and  $\psi_3$, which are given by $m_{\eta_{c2}(^1D_2)}=3.811$
 GeV and $m_{\psi_3(^3D_3)}=3.815$ GeV \cite{Ebert:2002pp}. { We vary $\pm50$ MeV to account for the uncertainties of these predicted masses. 
When taking $\alpha=0.70\pm0.05$, the same value }as that for $B^-\to X(3823) K^-$, we obtain the branching fractions for the processes $B^-\to \eta_{c2}K^-$ and $B^-\to \psi_3 K^-$,{
 \begin{eqnarray} \label{BRBtoX3etac2K}
 &&\mathcal{BR}(B^-\to \eta_{c2}(^1D_2) K^-)= (1.72\pm0.47)\times 10^{-5} \;,  \\
  &&\mathcal{BR}(B^-\to \psi_3(^3D_3) K^-)=( 0.80\pm0.21)\times 10^{-5} \; ,
 \end{eqnarray}
where the errors come from the uncertainties of $\alpha$ and the masses of $\eta_{c2}$ and  $\psi_3$.} The results are sizable and are the same order of magnitude as $B^-\to X(3823) K^-$, which means that these two decay channels can be accessible in future experiments. 

{There are some remarks on our theoretical uncertainties. The uncertainties come from three parts, the lack of real parts of the amplitudes, 
 the weak vertexes and the strong vertexes in loops. As for the real parts of the amplitudes, we assume they are not dominant as in Ref.~\cite{Cheng:2004ru}.
 As for the weak vertexes, there are actually much smaller uncertainties since either the naive factorization
 assumption for $B\to D^{(*)}\bar D_s^{(*)}$ (Eq.~(\ref{BtoDD})) or the form factor of the matrix element (Eq.~(\ref{IWfunction})) has been proven to
 have a good agreement with experiment. The dominant uncertainties come from strong vertexes: the coupling constants, the cutoff parameter $\alpha$ and the predicted masses
 of $\eta_{c2}(^1D_2)$ and $\psi_3(^3D_3)$. Since the coupling constants $g_H$ in Eq.~(\ref{lagrangianHH0-}) and $g_X$ in Eq.~(\ref{lagrangianXHH}) appear in
 all the amplitudes as global factors, after fitting to the process $B^-\to X(3823) K^-$, the uncertainties caused by $g_H$ and $g_X$ are just canceled when calculating 
 the processes $B^-\to \eta_{c2}(^1D_2) K^-$ and $B^-\to \psi_3(^3D_3) K^-$. The uncertainties from cutoff $\alpha$ and the predicted masses
 of $\eta_{c2}(^1D_2)$ and $\psi_3(^3D_3)$ have already considered in the text. As a whole, we stress that these uncertainties do not influence our main conclusion.}

\section{conclusions and discussion}\label{sec4}

The rescattering mechanism has been widely applied to the studies involved in hadronic transitions
 \cite{Liu:2006df,Meng:2007cx,Meng:2007tk,Liu:2009iw,Guo:2010zk,Chen:2011qx,Chen:2011zv,Guo:2016iej} and $B$ decays
  \cite{Colangelo:2002mj,Colangelo:2003sa,Cheng:2004ru,Wang:2008hq,Li:1996cj,Dai:1999cs}. As a long distant
   contribution, the rescattering mechanism is a typical non-perturbative QCD effect. 
Stimulated by the observation of $B^-\to X(3823) K^-$ \cite{Bhardwaj:2013rmw}, we study the contribution from the rescattering mechanism to 
$B^-\to X(3823) K^-$ since the naive factorization contribution to $B^-\to X(3823) K^-$ vanishes. With a reasonable cutoff
parameter, we can reproduce the experimental branching ratio of $B^-\to X(3823) K^-$. 
Under the same theoretical framework and with fitted parameters, we further investigate the processes
$B^-\to \eta_{c2}(^1D_2) K^-$ and $B^-\to \psi_3(^3D_3) K^-$. Our results show 
$\mathcal{BR}(B^-\to \eta_{c2}(^1D_2) K^-)= 1.7\times 10^{-5}$ and $\mathcal{BR}(B^-\to \psi_3(^3D_3) K^-)= 0.8\times 10^{-5}$, which are
comparable to $\mathcal{BR}(B^-\to X(3823) K^-)\simeq 2.1\times 10^{-5}$ extracted from experimental data. Our study shows that non-factorizable contribution to $B^-\to X(3823) K^-$, $B^-\to \eta_{c2}(^1D_2) K^-$ and $B^-\to \psi_3(^3D_3) K^-$ are sizable. 
Thus, experimental exploration of $B^-\to \eta_{c2}(^1D_2) K^-$ and $B^-\to \psi_3(^3D_3) K^-$ becomes possible at future experiments like LHCb, Belle, and the forthcoming BelleII. We also expect that our predictions of $B^-\to \eta_{c2}(^1D_2) K^-$ and $B^-\to \psi_3(^3D_3) K^-$ can be confirmed in experiments. 

{So far, two $D$-wave charmonia $\eta_{c2}(^1D_2)$
and $\psi_3(^3D_3)$ are still missing in experiments.
When exploring $B^-\to \eta_{c2}(^1D_2) K^-$ and $B^-\to \psi_3(^3D_3) K^-$, a key point  is how to identify $\eta_{c2}(^1D_2)$
and $\psi_3(^3D_3)$ experimentally, whose task is full of challenges faced by experimentalists. 
Since the present study shows that $B^-\to \eta_{c2}(^1D_2) K^-$ and $B^-\to \psi_3(^3D_3) K^-$ have sizable branching ratios. 
These two processes are also ideal channels to search for $\eta_{c2}(^1D_2)$
and $\psi_3(^3D_3)$. If future experiment can find these predicted decays, it
will not only make our knowledge of $B$ meson decays become more abundant, but also be helpful in establishing the 
charmonium family.} 

In summary, experimental study of $B^-\to \eta_{c2}(^1D_2) K^-$ and $B^-\to \psi_3(^3D_3) K^-$ will be a potential issue in near future. 
If these channels can be confirmed in experiments, the role of the rescattering mechanism in $B^-\to \eta_{c2}(^1D_2) K^-$ and $B^-\to \psi_3(^3D_3) K^-$
can be further identified, which will deepen our understanding of non-perturbative QCD behavior.

\section*{Acknowledgments}

This project is supported by the National Natural Science Foundation
of China under Grant No. 11222547 and No. 11175073. Xiang Liu is also supported by the Fundamental Research Funds for the Central Universities and the National Youth
Top-notch Talent Support Program ("Thousands-of-Talents Scheme").

\section*{Appendix: The rest of the amplitudes of the processes $B^-\to X(3823) K^-$, $B^-\to \eta_{c2} K^-$ and $B^-\to \psi_3 K^-$}
The amplitudes of the process $B^-\to X(3823) K^-$ depicted in the diagrams Fig.~\ref{hadronloopsof3823}~(1b)-(3b) are:
\begin{eqnarray}
&& Abs^{B^-\to\psi_2 K^-}_{(1b)}  \nonumber \\
&&= \frac{|p_2|}{32 \pi^2 m_1} \int d\Omega \frac{G_F}{\sqrt{2}} V_{cb} V^*_{cs} a_1 \sqrt{m_1 m_2} \xi(\omega) \left(v_1^\gamma + \frac{p_2^\gamma}{m_2} \right)
f_3 p_{3\gamma}  \nonumber \\
&&\quad \times  (-1) g_{DD_s^*K}  p^\alpha_6  \left( -g_{\alpha\mu} + \frac{p_{4\alpha}p_{4\mu}}{m^2_4} \right) (-i) g_{\psi_2D_sD^*_s} \epsilon^{*\mu\nu}_5 
\nonumber \\
&& \quad\times (p_{3\nu}-p_{4\nu}) \frac{1}{p^2_4-m^2_4} \mathfrak{F}^2(p_4^2) \; ,
\end{eqnarray}
\begin{eqnarray}
&& Abs^{B^-\to\psi_2 K^-}_{(2a)}  \nonumber \\
&&= \frac{|p_2|}{32 \pi^2 m_1} \int d\Omega \frac{G_F}{\sqrt{2}} V_{cb} V^*_{cs} a_1 \sqrt{m_1 m_2} \xi(\omega)
\Bigg( i\varepsilon_{\gamma\delta\alpha\beta} \frac{p^\alpha_2}{m_2}v^\beta_1  \nonumber \\
&& \quad- (1+\omega)g_{\delta\gamma} + v_{1\delta} \frac{p_{2\gamma}}{m_2} \Bigg) f_3 m_3 \left( -g^\delta_\mu + \frac{p^\delta_2 p_{2\mu}}{m^2_2} \right)
i g_{\psi_2DD^*} \epsilon^{*\mu\nu}_5  \nonumber \\
&&\quad \times (p_{4\nu}+p_{2\nu}) (-g_{DD^*_sK}) p_{6\theta} \left( -g^{\theta\gamma} + \frac{p^\theta_3 p^\gamma_3}{m^2_3} \right)
\frac{1}{p^2_4-m^2_4}  \nonumber \\
&&\quad \times \mathfrak{F}^2(p^2_4)  \; ,
\end{eqnarray}
\begin{eqnarray}
&& Abs^{B^-\to\psi_2 K^-}_{(2b)}  \nonumber \\
&&= \frac{|p_2|}{32 \pi^2 m_1} \int d\Omega \frac{G_F}{\sqrt{2}} V_{cb} V^*_{cs} a_1 \sqrt{m_1 m_2} \xi(\omega)
\Bigg( i\varepsilon_{\gamma\delta\alpha\beta} \frac{p^\alpha_2}{m_2}v^\beta_1  \nonumber \\
&& \quad- (1+\omega)g_{\delta\gamma} + v_{1\delta} \frac{p_{2\gamma}}{m_2} \Bigg) f_3 m_3 \left( -g^\delta_\theta + \frac{p^\delta_2 p_{2\theta}}{m^2_2} \right)
g_{D_sD^*K} p^\theta_6 (-i) \nonumber \\
&& \quad\times g_{\psi_2D_sD^*_s} \epsilon^{*\mu\nu}_5 (p_{4\nu}-p_{3\nu}) \left( -g^\gamma_\mu + \frac{p_{3\mu}p^\gamma_3}{m^2_3} \right)
\frac{1}{p^2_4-m^2_4}  \nonumber \\
&& \quad\times \mathfrak{F}^2(p^2_4)  \; ,
\end{eqnarray}
\begin{eqnarray}
&& Abs^{B^-\to\psi_2 K^-}_{(3a)}  \nonumber \\
&&= \frac{|p_2|}{32 \pi^2 m_1} \int d\Omega \frac{G_F}{\sqrt{2}} V_{cb} V^*_{cs} a_1 \sqrt{m_1 m_2} \xi(\omega)
\Bigg( i\varepsilon_{\gamma\delta\zeta\eta} \frac{p^\zeta_2}{m_2}v^\eta_1  \nonumber \\
&& \quad- (1+\omega)g_{\delta\gamma} + v_{1\delta} \frac{p_{2\gamma}}{m_2} \Bigg) f_3 m_3 (-i) g_{\psi_2D^*D^*}
\Bigg[ \left( -g^\delta_\nu + \frac{p^\delta_2 p_{2\nu}}{m^2_2} \right) \nonumber \\
&& \quad\times \varepsilon_{\mu\rho\alpha\beta} \epsilon^{*\mu\nu}_5 p^\alpha_4 p^\beta_5 \left( -g^{\rho\sigma} + \frac{p^\rho_4 p^\sigma_4}{m^2_4} \right) 
+ \left( -g^{\delta\rho} + \frac{p^\delta_2 p^\rho_2}{m^2_2} \right) \varepsilon_{\mu\rho\alpha\beta} \nonumber \\
&&\quad \times \epsilon^{*\mu\nu}_5 p^\alpha_4 p^\beta_5 \left( -g^\sigma_\nu + \frac{p_{4\nu} p^\sigma_4}{m^2_4} \right) \Bigg]
g_{D^*D^*_sK} \varepsilon_{\tau\sigma\kappa\lambda} p^\kappa_4 p^\lambda_6 \nonumber \\
&&\quad \times \left( -g^{\tau\gamma} + \frac{p^\tau_3 p^\gamma_3}{m^2_3} \right) \frac{1}{p^2_4-m^2_4} \mathfrak{F}^2(p^2_4) \;,
\end{eqnarray}
\begin{eqnarray}
&& Abs^{B^-\to\psi_2 K^-}_{(3b)}  \nonumber \\
&&= \frac{|p_2|}{32 \pi^2 m_1} \int d\Omega \frac{G_F}{\sqrt{2}} V_{cb} V^*_{cs} a_1 \sqrt{m_1 m_2} \xi(\omega)
\Bigg( i\varepsilon_{\gamma\delta\zeta\eta} \frac{p^\zeta_2}{m_2}v^\eta_1  \nonumber \\
&&\quad - (1+\omega)g_{\delta\gamma} + v_{1\delta} \frac{p_{2\gamma}}{m_2} \Bigg) f_3 m_3 \left( -g^{\delta\theta} + \frac{p^\delta_2 p^\theta_2}{m^2_2} \right)
g_{D^*D^*_sK} \varepsilon_{\sigma\theta\tau\kappa} \nonumber \\
&&\quad\times p^\tau_2 p^\kappa_6 i g_{\psi_2D^*_sD^*_s} \Bigg[ \left( -g^\sigma_\nu + \frac{p^\sigma_4 p_{4\nu}}{m^2_4} \right) 
\varepsilon_{\mu\rho\alpha\beta} \epsilon^{*\mu\nu}_5 p^\alpha_3 p^\beta_5 \nonumber \\
&& \quad\times \left( -g^{\rho\gamma} + \frac{p^\rho_3 p^\gamma_3}{m^2_3} \right) + \left( -g^{\sigma\rho} + \frac{p^\sigma_4 p^\rho_4}{m^2_4} \right)
\varepsilon_{\mu\rho\alpha\beta} \epsilon^{*\mu\nu}_5 p^\alpha_3 p^\beta_5 \nonumber \\
&& \quad\times \left( -g^\gamma_\nu + \frac{p_{3\nu} p^\gamma_3}{m^2_3} \right) \Bigg] \frac{1}{p^2_4-m^2_4} \mathfrak{F}^2(p^2_4) \;.
\end{eqnarray}

The amplitudes of the process $B^-\to \eta_{c2} K^-$ depicted in the diagrams Fig.~\ref{hadronloopsof3823}~(1b)-(3b) are:
\begin{eqnarray}
&& Abs^{B^-\to \eta_{c2} K^-}_{(1b)}  \nonumber \\
&&= \frac{|p_2|}{32 \pi^2 m_1} \int d\Omega \frac{G_F}{\sqrt{2}} V_{cb} V^*_{cs} a_1 \sqrt{m_1 m_2} \xi(\omega) \left(v_1^\gamma + \frac{p_2^\gamma}{m_2} \right)
f_3 p_{3\gamma}  \nonumber \\
&&\quad \times  (-1) g_{DD_s^*K}  p^\alpha_6  \left( -g_{\alpha\mu} + \frac{p_{4\alpha}p_{4\mu}}{m^2_4} \right)  g_{\eta_{c2}D_sD^*_s} \epsilon^{*\mu\nu}_5 
(p_{3\nu}-p_{4\nu}) \nonumber \\
&&\quad \times  \frac{1}{p^2_4-m^2_4} \mathfrak{F}^2(p_4^2) \; ,
\end{eqnarray}
\begin{eqnarray}
&& Abs^{B^-\to \eta_{c2} K^-}_{(2a)}  \nonumber \\
&&= \frac{|p_2|}{32 \pi^2 m_1} \int d\Omega \frac{G_F}{\sqrt{2}} V_{cb} V^*_{cs} a_1 \sqrt{m_1 m_2} \xi(\omega)
\Bigg( i\varepsilon_{\gamma\delta\alpha\beta} \frac{p^\alpha_2}{m_2}v^\beta_1  \nonumber \\
&& \quad- (1+\omega)g_{\delta\gamma} + v_{1\delta} \frac{p_{2\gamma}}{m_2} \Bigg) f_3 m_3 \left( -g^\delta_\mu + \frac{p^\delta_2 p_{2\mu}}{m^2_2} \right)
(-1) g_{\eta_{c2}DD^*}   \nonumber \\
&& \quad\times \epsilon^{*\mu\nu}_5 (p_{4\nu}+p_{2\nu}) (-g_{DD^*_sK}) p_{6\theta} \left( -g^{\theta\gamma} + \frac{p^\theta_3 p^\gamma_3}{m^2_3} \right)
\frac{1}{p^2_4-m^2_4}  \nonumber \\
&&\quad \times \mathfrak{F}^2(p^2_4)  \; ,
\end{eqnarray}
\begin{eqnarray}
&& Abs^{B^-\to \eta_{c2} K^-}_{(2b)}  \nonumber \\
&&= \frac{|p_2|}{32 \pi^2 m_1} \int d\Omega \frac{G_F}{\sqrt{2}} V_{cb} V^*_{cs} a_1 \sqrt{m_1 m_2} \xi(\omega)
\Bigg( i\varepsilon_{\gamma\delta\alpha\beta} \frac{p^\alpha_2}{m_2}v^\beta_1  \nonumber \\
&&\quad - (1+\omega)g_{\delta\gamma} + v_{1\delta} \frac{p_{2\gamma}}{m_2} \Bigg) f_3 m_3 \left( -g^\delta_\theta + \frac{p^\delta_2 p_{2\theta}}{m^2_2} \right)
g_{D_sD^*K} p^\theta_6 (-1) \nonumber \\
&&\quad \times g_{\eta_{c2}D_sD^*_s} \epsilon^{*\mu\nu}_5 (p_{4\nu}-p_{3\nu}) \left( -g^\gamma_\mu + \frac{p_{3\mu}p^\gamma_3}{m^2_3} \right)
\frac{1}{p^2_4-m^2_4}   \nonumber \\
&&\quad \times \mathfrak{F}^2(p^2_4)   \; ,
\end{eqnarray}
\begin{eqnarray}
&& Abs^{B^-\to \eta_{c2} K^-}_{(3a)}  \nonumber \\
&&= \frac{|p_2|}{32 \pi^2 m_1} \int d\Omega \frac{G_F}{\sqrt{2}} V_{cb} V^*_{cs} a_1 \sqrt{m_1 m_2} \xi(\omega)
\Bigg( i\varepsilon_{\gamma\delta\zeta\eta} \frac{p^\zeta_2}{m_2}v^\eta_1  \nonumber \\
&&\quad - (1+\omega)g_{\delta\gamma} + v_{1\delta} \frac{p_{2\gamma}}{m_2} \Bigg) f_3 m_3  \left( -g^{\delta\alpha} + \frac{p^\delta_2 p^\alpha_{2}}{m^2_2} \right)
(-1) g_{\eta_{c2}D^*D^*} \nonumber \\
&&\quad \times \varepsilon_{\rho\nu\alpha\beta} p^\rho_5 \epsilon^{*\mu\nu}_5 p_{4\mu} \left( -g^{\beta\sigma} + \frac{p^\beta_4 p^\sigma_4}{m^2_4} \right) 
g_{D^*D^*_sK} \varepsilon_{\tau\sigma\kappa\lambda} p^\kappa_4 p^\lambda_6 \nonumber \\
&&\quad \times \left( -g^{\tau\gamma} + \frac{p^\tau_3 p^\gamma_3}{m^2_3} \right) \frac{1}{p^2_4-m^2_4} \mathfrak{F}^2(p^2_4) \;,
\end{eqnarray}
\begin{eqnarray}
&& Abs^{B^-\to \eta_{c2} K^-}_{(3b)}  \nonumber \\
&&= \frac{|p_2|}{32 \pi^2 m_1} \int d\Omega \frac{G_F}{\sqrt{2}} V_{cb} V^*_{cs} a_1 \sqrt{m_1 m_2} \xi(\omega)
\Bigg( i\varepsilon_{\gamma\delta\zeta\eta} \frac{p^\zeta_2}{m_2}v^\eta_1  \nonumber \\
&& \quad- (1+\omega)g_{\delta\gamma} + v_{1\delta} \frac{p_{2\gamma}}{m_2} \Bigg) f_3 m_3 \left( -g^{\delta\theta} + \frac{p^\delta_2 p^\theta_2}{m^2_2} \right)
g_{D^*D^*_sK} \varepsilon_{\sigma\theta\tau\kappa} \nonumber \\
&&\quad\times p^\tau_2 p^\kappa_6 \left( -g^{\sigma\alpha} + \frac{p^\sigma_4 p^{\alpha}_{4}}{m^2_4} \right)  g_{\eta_{c2}D^*_sD^*_s}  
\varepsilon_{\rho\nu\alpha\beta} p^\rho_5 \epsilon^{*\mu\nu}_5 p_{3\mu}  \nonumber \\
&&\quad \times \left( -g^{\beta\gamma} + \frac{p^\beta_3 p^\gamma_3}{m^2_3} \right)  \frac{1}{p^2_4-m^2_4} \mathfrak{F}^2(p^2_4) \;.
\end{eqnarray}

The amplitudes of the process $B^-\to \psi_3 K^-$ depicted in the diagrams Fig.~\ref{FeynmannDiagramX3}~(1b)-(2b) are:
\begin{eqnarray}
&& Abs^{B^-\to \psi_3 K^-}_{(1b)}  \nonumber \\
&&= \frac{|p_2|}{32 \pi^2 m_1} \int d\Omega \frac{G_F}{\sqrt{2}} V_{cb} V^*_{cs} a_1 \sqrt{m_1 m_2} \xi(\omega)
\Bigg( \frac{p^\beta_2}{m_2}+v^\beta_1  \Bigg) f_3 m_3 \nonumber \\
&&\quad (-1) g_{DD^*_sK} p^\theta_6  \left( -g_{\theta\alpha} + \frac{ p_{4\theta} p_{4\alpha}}{m^2_4} \right) i g_{\psi_3D^*_sD^*_s} 
\epsilon^{*\mu\nu\alpha}_5  (p_{4\mu}-p_{3\mu}) \nonumber \\
&& \quad\times   \left( -g_{\nu\beta} + \frac{p_{3\nu}p_{3\beta}}{m^2_3} \right)
\frac{1}{p^2_4-m^2_4}  \mathfrak{F}^2(p^2_4)   \; , 
\end{eqnarray}
\begin{eqnarray}
&& Abs^{B^-\to \psi_3 K^-}_{(2a)}  \nonumber \\
&&= \frac{|p_2|}{32 \pi^2 m_1} \int d\Omega \frac{G_F}{\sqrt{2}} V_{cb} V^*_{cs} a_1 \sqrt{m_1 m_2} \xi(\omega)
\Bigg( i\varepsilon_{\gamma\delta\zeta\eta} \frac{p^\zeta_2}{m_2}v^\eta_1  \nonumber \\
&& \quad- (1+\omega)g_{\delta\gamma} + v_{1\delta} \frac{p_{2\gamma}}{m_2} \Bigg) f_3 m_3  \left( -g^{\delta}_{\alpha} + \frac{p^\delta_2 p_{2\alpha}}{m^2_2} \right)
i g_{\psi_3D^*D^*} \epsilon^{*\mu\nu\alpha}_5\nonumber \\
&& \quad\times  (p_{4\mu}+p_{2\mu}) \left( -g^{\tau}_{\nu} + \frac{p_{4\nu} p^\tau_4}{m^2_4} \right) 
g_{D^*D^*_sK} \varepsilon_{\sigma\tau\rho\beta} p^\rho_4 p^\beta_6 \nonumber \\
&&\quad \times \left( -g^{\sigma\gamma} + \frac{p^\sigma_3 p^\gamma_3}{m^2_3} \right) \frac{1}{p^2_4-m^2_4} \mathfrak{F}^2(p^2_4) \;,
\end{eqnarray}
\begin{eqnarray}
&& Abs^{B^-\to \psi_3 K^-}_{(2b)}  \nonumber \\
&&= \frac{|p_2|}{32 \pi^2 m_1} \int d\Omega \frac{G_F}{\sqrt{2}} V_{cb} V^*_{cs} a_1 \sqrt{m_1 m_2} \xi(\omega)
\Bigg( i\varepsilon_{\gamma\delta\zeta\eta} \frac{p^\zeta_2}{m_2}v^\eta_1  \nonumber \\
&&\quad - (1+\omega)g_{\delta\gamma} + v_{1\delta} \frac{p_{2\gamma}}{m_2} \Bigg) f_3 m_3 \left( -g^{\delta\tau} + \frac{p^\delta_2 p^\tau_2}{m^2_2} \right)
g_{D^*D^*_sK} \varepsilon_{\sigma\tau\rho\beta} \nonumber \\
&&\quad\times p^\rho_2 p^\beta_6 \left( -g^{\sigma}_{\alpha} + \frac{p^\sigma_4 p_{4\alpha}}{m^2_4} \right) i g_{\psi_3D^*_sD^*_s}  
\epsilon^{*\mu\nu\alpha}_5 (p_{4\mu}-p_{3\mu})  \nonumber \\
&& \quad\times \left( -g^{\gamma}_\nu + \frac{p_{3\nu} p^\gamma_3}{m^2_3} \right)  \frac{1}{p^2_4-m^2_4} \mathfrak{F}^2(p^2_4) \;.
\end{eqnarray}

\end{document}